\documentclass[usenatbib]{mn2e}
\usepackage{natbib}
\usepackage{amsmath}
\usepackage{multirow}
\usepackage{array}
\usepackage{url} 
\usepackage{epsfig}
\usepackage{verbatim} 
\usepackage[usenames]{color}
\allowdisplaybreaks[1]

\definecolor{purple}{RGB}{160,32,240}

\newcommand{\peter}[1]{}

\newcommand{\Msun}{{\mathrm M}_{\odot}}
\newcommand{\Msol}{{\mathrm M}_{\odot}}

\newcommand{\erg}{{\rm erg}}
\newcommand{\s}{{\rm s}}

\newcommand{\apj}{ApJ}
\newcommand{\apjl}{ApJ}
\newcommand{\apjs}{ApJS}
\newcommand{\aap}{A$\&$A}
\newcommand{\aapr}{A$\&$A}

\newcommand{\mnras}{MNRAS}

\newcommand{\prd}{PRD}
\newcommand{\aj}{AJ}
\newcommand{\nat}{Nature}

  \def\spose#1{\hbox to 0pt{#1\hss}}
  \def\ltsim{\mathrel{\spose{\lower.5ex\hbox{$\mathchar"218$}}
  \raise.4ex\hbox{$\mathchar"13C$}}}

\defcitealias{tph12}{TPH12}
\defcitealias{TH09}{TH09}

\title[Baryonic streaming motions and SMBH formation]{The effect of baryonic streaming motions on the formation of the first supermassive black holes}

\author[T.L.Tanaka, M.Li and Z.Haiman]
{Takamitsu L. Tanaka$^{1}$\thanks{E-mail: taka@mpa-garching.mpg.de}, Miao Li$^{2}$ and Zolt\'an Haiman$^{2}$\\
$^{1}$Max Planck Institute for Astrophysics, Karl-Schwarzschild-Str. 1, D-85741 Garching, Germany\\
$^{2}$Department of Astronomy, Columbia University, 550 W. 120th Street, New York, NY 10027, USA}
\begin{document}

\maketitle
\label{firstpage}

\begin{abstract}
Observations of quasars at redshifts $z\ga 6$ reveal
that $10^9 \,\Msol$ supermassive black holes (SMBHs)
had already formed when the Universe was $\ltsim 0.9\,\mathrm{Gyr}$ old.
One hypothesis for the origins of these SMBHs is that they grew
from the remnants of the first generation of massive stars,
which formed in low-mass ($\sim10^{5-6}\,\Msol$) dark matter minihaloes at $z\ga 20$.
This is the regime where baryonic streaming motions---the
relative velocities of baryons with respect to dark matter in the early Universe---most
strongly inhibit star formation by suppressing gas infall and cooling. %
We investigate the impact of this effect on the growth of the first SMBHs
using a suite of high-fidelity, ellipsoidal-collapse Monte Carlo merger-tree simulations.
We find that the suppression of seed BH formation by the streaming motions
significantly reduces the number density of the most massive BHs at $z>15$,
but the residual effect at lower redshifts is essentially negligible.
The streaming motions can reduce by a factor of few
the number density of the most luminous quasars at $z\approx 10-11$,
where such objects could be detected
by the {\it James Webb Space Telescope}.
We conclude, with minor theoretical caveats,
that baryonic streaming motions are unlikely to pose
a significant additional obstacle to the formation of the observed high-redshift quasar SMBHs.
Nor do they appreciably affect the heating and reionization histories of the Universe
or the merger rates of nuclear BHs in the mass and redshift ranges
of interest for proposed gravitational-wave detectors.
\end{abstract}


\section{Introduction}
\label{sec:intro}

Recent surveys such as SDSS, CFHQS and UKIDSS have unveiled about 50 quasars at redshifts $z\ga 6$, when the Universe
is less than 1 Gyr old
\citep{fan01,fan06,lawrence07,willott09}.
The most distant quasar discovered to date, ULAS J1120+0641, 
has a redshift of $z=7.1$ ($\approx 0.8~{\rm Gyr}$ after the big bang),
with a mass $\sim 2\times 10^{9}~\Msol$ \citep{mortlock11}.

It is still a mystery how these supermassive black holes (SMBHs)
accumulated so much mass in such a short time
 (see reviews by \citealt{volonteri10} and \citealt{Haiman13}).
 One possibility is that they grew from BHs left behind by the first generation of stars
(e.g. \citealt{MadauRees01}, \citealt{HaimanLoeb01},
\citealt{Bromley+04}, \citealt{Shapiro05}, \citealt{Pelupessy+07}, \citealt{TH09}, hereafter TH09).
These `Population III' (PopIII) stars are thought to have formed from molecular-cooling gas collapsing inside dark matter (DM) minihaloes at redshifts $z\ga20$ \citep[e.g.][]{bl04}.
PopIII stars are thought to be
massive ($\sim 30-300~\Msol$, \citealt{Heger+03}, \citealt{Ohkubo+09};
but see \citealt{Turk+09}, \citealt{Stacy+10}, \citealt{Hosokawa+11}, \citealt{Greif+11b}),
metal-free, and short-lived;
after their deaths, they would have left behind BHs 
with $\sim 40$ per cent of their original mass \citep[e.g.][]{ZhangW+08}.
These `seed' BHs can then grow by accreting gas
and merging with their peers through hierarchical structure formation.
Provided that they can accrete efficiently (but see \citealt{Milos+09}, \citealt{Alvarez+09})
PopIII seeds can  grow to $\sim 10^{9}\Msun$ 
by $z\approx7$ without a super-Eddington phase \citep[e.g.][hereafter TPH12]{tph12}.

One of the main uncertainties of the  above scenario of SMBH formation
is when and where the first stars form.
Recent studies of the  effect of baryonic streaming motions (BSMs)
have put new insights into this question \citep[e.g.][]{th10}.
Before cosmic recombination, the excitation of acoustic oscillations in
photon-baryon fluids generates relative bulk velocities between the gas and the DM.
During recombination, the sound speed of baryons drops abruptly,
so that the mean relative motion of $\sim 30 ~{\rm km/s}$ becomes supersonic.
This bulk motion makes it easier for gas to stream out of
DM haloes and thus affects the distribution and evolution of baryons in the early Universe
\citep{Dalal+10,Maio+11, tbh11,fbth12}.
In particular, the streaming velocities suppress the formation
of PopIII stars \citep{Greif+11a,sbl11}.
The effect is larger at higher redshifts, when the characteristic
DM halo potentials are shallower and the bulk streaming velocities
are larger compared to the sound speed of the intergalactic medium (IGM). 

The earliest-forming PopIII seeds,
which are the most vulnerable to BSMs,
are also believed to be the most important `building blocks'
in the assembly of the first SMBHs in the Pop-III scenario.
This is because of the gravitational recoil effect, in which the asymmetric
emission of gravitational waves imparts velocities
as large as thousands of ${\rm km~s}^{-1}$ to the merged object,
relative to the rest frame of the binary's centre of mass
(\citealt{Peres62}, \citealt{Kidder95}, \citealt{Favata+04}, \citealt{Baker+06b}).
Mergers of BHs of similar masses result in higher recoil velocities,
and are thus more likely to result in ejection from the host halo (\citealt{Haiman04}, \citealt{YooME04}).
Inversely, a seed that has formed with a greater mass than the rest of
the population will suffer smaller recoils following a merger
and is less likely to be ejected;
a seed that forms earlier than the rest may be doubly protected,
because it will have had more time to grow before its first merger
and also benefit from residing in a deeper host potential well.
The survival bias for the most massive BHs
(`survival of the fittest'; \citealt{VolRees06}; cf. \citetalias{TH09})
increases in a runaway fashion as the objects that
survive mergers become increasingly more massive
with respect to their contemporaries.
BSMs could thus affect the assembly of SMBHs via mergers by
preferentially suppressing the number density of the earliest seed BHs.

In this work, we investigate how this reduction of the number of seed BHs affects
SMBH formation in the early Universe. In particular, we are interested in the following two questions: 
(1) whether the formation of $10^{9}~\Msun$ SMBHs at $z\geq 7$ is still possible without super-Eddington accretion;
and (2) how the BSMs influence the BH mass function in the redshift range $z=6-11$, which could be probed by the \textit{James Webb Space Telescope}\footnote{
\url{http://www.jwst.nasa.gov/}} (\textit{JWST}).

We organize our paper as follows.
In \S\ref{s:Methods}, we describe our semianalytic model:
the construction of merger trees based on a new algorithm,
the implementation of the BSM effect,
prescriptions for BH formation and growth,
and the subsequent heating of the IGM.
We present and discuss our results in \S\ref{s:results},
and conclude in \S\ref{s:concl}.

Throughout this paper, we adopt a $\Lambda$ cold dark matter ($\Lambda$CDM) cosmology with
parameters from seven-year results of the \textit{Wilkinson Microwave Anisotropy Probe}
(\textit{WMAP}7; \citealt{Jarosik+11}):
$\Omega_{\rm CDM} = 0.227$, $\Omega_{\rm b} = 0.0455$,
$\Omega_{\Lambda} = 0.728$,
$h=0.702$, $n_{\rm s} = 0.961$, and $\sigma_{8} = 0.807$.
These parameters are consistent with the \textit{WMAP} nine-year results \citep{Hinshaw+12}
as well as the \textit{Planck} results \citep{PlanckParameters}
within the $1$-$\sigma$ uncertainties.
The variance of density fluctuations is calculated by using fitting formulae for the DM power spectrum
\citep[from][]{eh98} and the growth function
\citep[from][]{cpt92}.

\section{Methods}
\label{s:Methods}

\subsection{The Merger Tree}
Following previous works (\citealt{VHM03}, \citealt{YooME04}, \citealt{Bromley+04}, \citetalias{TH09}),
we employ semianalytic Monte Carlo merger trees to simulate
the hierarchical assembly of DM haloes and the growth of their nuclear BHs. An important aspect of the present work is the adoption of the ellipsoidal 
collapse model \citep{st02} of DM in the extended Press-Schechter (EPS) formalism.
 Previous studies that utilized  merger trees have used the spherical collapse model \citep{LaceyCole93},
which is known to underpredict the number of the most massive haloes when compared to $N$-body simulations,
 with the discrepancy increasing with look-back time.
 
However, until recently it was not practically feasible
to generate merger trees based on the ellipsoidal EPS model,
because the fitting formula for the conditional mass function 
was inaccurate for small time steps $\Delta z$.
\citet{zmf08} derived an accurate form of the conditional mass function
for $\Delta z\ltsim 0.1$---thus enabling the construction of ellipsoidal merger
trees---and developed several algorithms that faithfully reproduce the progenitor mass function.
We adopt their `method B', which in a follow-up paper (Zhang et al. in preparation)
is found to agree best with $N$-body simulations.
We do not detail the mathematical formulae and algorithms here,
and instead refer the reader to the above papers for a comprehensive description.

We have confirmed the fidelity of our merger trees by
comparing the progenitor mass function with the semianalytic predictions
of the ellipsoidal EPS model. For example, in 100 realizations of the assembly history of a
 $10^{12}~\Msun$ parent halo at $z=6$,
 the mean Monte Carlo progenitor mass function agrees with the theoretical expectation
to within a few percent out to $z\sim 30$.
The discrepancy becomes somewhat larger for higher redshift
when there are fewer progenitors, e.g. within $\sim 30$ per cent for $z=40$.
We note that compared to the spherical-collapse model,
the ellipsoidal model predicts at $z\sim 40$ a factor $\sim 70$ more 
haloes with $M> 6\times10^{5}\Msun$.
Thus, employing the latter model is essential in accurately
characterizing the population of the very first seed BHs.

We simulate the assembly history of $70$ haloes whose masses $M_{\rm halo}$
at $z=6$ exceed $10^{12.85}~\Msol$;
these haloes represent a comoving volume of $\sim 150~{\rm Gpc}^3$.
Because it is intractable to directly simulate all of the haloes
below $10^{12.85}~\Msol$ in such a large volume,
we instead construct a statistical representation of these lower-mass haloes
by simulating narrow mass bins---with logarithmic widths of
$\Delta \log M_{\rm halo}\le 0.5$ ranging
from $\log (M_{\rm halo}/\Msol)=8$-$12.85$---each
represented by a sample of 100 haloes, and then scaling up the counting results for each bin
to the expected halo abundances (see section 2.6 and table 1 in  \citetalias{TH09} for details).
In other words, our simulations combine a statistical approximation
of the assembly of haloes with $10^8~\Msol <M_{\rm halo}(z=6)<10^{12.85}~\Msol$,
alongside a direct realization of those with $M_{\rm halo}(z=6)>10^{12.85}~\Msol$.

\subsection{Implementation of Baryonic Streaming Motions}
BSMs suppress the formation of the first stars by inhibiting
the collapse of gas into early DM minihaloes.
Several numerical studies have shown that in the absence of BSMs,
PopIII stars typically form in haloes with virial temperatures $T_{\rm vir}\sim 1000~{\rm K}$
\citep[e.g.][and references therein]{Bromm+09}
or, equivalently, circular velocities $v_{\rm circ}\sim 4~{\rm km~s}^{-1}$.
Streaming velocities effectively raise this threshold for PopIII formation;
\cite{fbth12} fit the following expression for the `cooling' circular velocity threshold
to the simulation results of \cite{sbl11} and \cite{Greif+11a}
\begin{equation}
v_{\rm cool}=\sqrt{v_0^{2}+\left[\alpha ~ v_{\rm BSM}(z)\right]^{2}},
\label{eq:vcool}
\end{equation}
with $v_0=3.7~{\rm km/s}$ and $\alpha=4.0$.
The case $v_{\rm BSM}=0$ corresponds to $T_{\rm vir}\approx 960~{\rm K}$.

Because BSMs are coherent on scales of a few comoving Mpc \citep{th10}
and the DM haloes in our simulation satisfy $(M_{\rm halo}/\rho_{\rm m})^{1/3} \ltsim {\rm Mpc}$,
we associate each $z=6$ parent halo and all of its progenitors with a single
streaming velocity value at recombination, $v_{\rm BSM}^{\rm (rec)}$,
drawn from a Maxwell-Boltzmann distribution
with an rms value $\sigma_{\rm BSM}^{\rm (rec)}\approx 30~{\rm km/s}$.
The streaming velocity subsequently decays with time as $v_{\rm BSM}(z)\propto (1+z)$. 
 
We convert the circular velocity threshold in equation \eqref{eq:vcool}
to a mass threshold $M_{\rm cool}\left(z, v_{\rm BSM}^{\rm (rec)}\right)$
through the relation $v_{\rm cool}=\sqrt{GM_{\rm cool}/r_{\rm vir}}$,
where $r_{\rm vir}$ is the virial radius;
\begin{equation}
M_{\rm cool}\approx 4.5\times 10^5 \left(\frac{v_{\rm cool}}{4~{\rm km}~\s^{-1}}\right)^3
\left(\frac{1+z}{21}\right)^{-3/2}\Msol.
\end{equation}
We seed a halo with the BH remnant of a PopIII star if its mass is higher
than both $M_{\rm cool}$ and the cosmological Jeans mass (e.g. \citealt{BarkanaLoeb01}):
\begin{equation}
\label{eq:MJ}
M_{\rm J} = \frac{4\pi}{3}\rho_0\left(\frac{\lambda_{\rm J}}{2}\right)^3,
\end{equation}
where $\lambda_{\rm J}=c_{\rm s}\sqrt{\pi/(G\rho_0)}$ is the Jeans length,
$\rho_0(z)=3H_0^2/(8\pi G)\times \Omega_0(1+z)^3$ is the mean matter density and
$c_{\rm s}=\sqrt{5k_{\rm B}T_{\rm IGM}/(3\mu m_{\rm p})}$ is the gas sound speed in the IGM.
(Here, $G$, $k_{\rm B}$ and $m_{\rm p}$ are the usual physical constants,
$H=100h ~{\rm km}~ \s^{-1}~ {\rm Mpc}^{-1}$ is the Hubble parameter,
$T_{\rm IGM}$ is the IGM temperature, and $\mu$ its mean molecular weight.)

In the absence of baryonic streaming, $M_{\rm J}\propto c_{\rm s}^{3} (1+z)^{-3/2}$.
However, with BSMs the effective Jeans length scales as
$\lambda_{\rm J}\propto \sqrt{c_{\rm s}^2+v_{\rm BSM}^2}$ \citep{sbl11},
so that 
\begin{equation}
\label{eq:equation}
M_{\rm J}\approx 5.6\times 10^5
\left(\frac{\sqrt{c_{\rm s}^2+v_{\rm BSM}^2}}{1~{\rm km}~{\rm s}^{-1}}\right)^{3}
\left(\frac{1+z}{21}\right)^{-3/2}\Msol.
\end{equation}
We compute the IGM temperature and the corresponding Jeans mass
self-consistently by calculating the photoheating due to the X-rays produced by accreting BHs (see below).
As a control, we have also run simulations with $v_{\rm BSM}$ set to zero,
in which $M_{\rm cool}$ and $M_{\rm J}$ have uniform values
at any given values of $z$ and $T_{\rm IGM}$.

\subsection{Other Baryonic Processes}
 We adopt several semi-analytic prescriptions to treat relevant
baryonic processes that contribute to SMBH growth.
Many of the model components listed below are described in greater detail in \citetalias{tph12},
to which we refer the reader for specifics.

\emph{BH seeding and IMF.}
The initial mass function (IMF) of the stellar population is highly uncertain;
recent simulations suggest that the typical mass maybe few tens of $\Msol$
\citep{Turk+09, Stacy+10, Greif+11b},
but some stars may form with much higher masses \citep{OmukaiPalla03, Ohkubo+09}.
We adopt a Salpeter IMF with a floor of $20~\Msol$ and slope ${\rm d}n/{\rm d}\log M_*\propto M_*^{-1.35}$.
The masses of the remnant seed BHs are prescribed using a semianalytic fit (\citetalias{tph12})
to the simulation results of \cite{ZhangW+08}.
For the high-mass stars ($M_*>45~\Msol$) of the greatest interest,
the remnant BH mass is $\sim 20-40$ per cent of the stellar mass;
we assume that stars in the pair-instability mass window $140~\Msol<M_*<260~\Msol$
leave behind no BH remnant \citep{Heger+03}.
We place a single seed BH in each DM halo that satisfies
a minimum-mass criterion, as described above.
A halo is not seeded if it or any of its progenitors have
previously formed a seed BH.

\emph{BH mergers and recoil.}
We compute the merger time of a merging DM halo using the fitting function
of \cite{BoylanKolchin+08}, assuming a circularity parameter $\eta=0.5$.
If the merger time is longer than the Hubble time, we assume that the secondary halo
ends up `stuck' as a satellite and its BH is for all practical purposes removed from
the simulation.
If the merger time is shorter than the Hubble time, the two nuclear BHs are
assumed to merge on the same time-scale as the host haloes (for the purposes
of the merger-tree simulation, instantaneously).
 A random gravitational recoil velocity is assigned to the merged BH
according to the fitting formula of \cite{lousto10},
under the assumption that components of BH spin are uniformly
distributed: $0<a<0.93$ for the dimensionless spin magnitude,
and $0<\theta<\pi/6$ for spin vector angles with respect to the inspiral plane \citep[see][]{Bogdanovic+07,Dotti+09}.
The recoiling BH is discarded if the recoil velocity is so great that
the BH cannot resettle at the halo centre via damping by dynamical friction
within a Hubble time (see \citetalias{TH09} for details).

\emph{BH growth and IGM heating.}
We assume a simple growth model in which
BHs are assumed to grow exponentially
at a mean rate of $2/3$ of the Eddington rate,
with a time-averaged radiative efficiency $\epsilon=0.07$---\textit{provided that a rich supply of cold gas is available} (see paragraph below).
We assume that growing BHs emit $90$ per cent of their radiation as a multicolour disc \citep{ss73}
with a greybody spectrum \citep{Blaes04,TM10},
and the other $10$ per cent as a hard X-ray corona above $1~{\rm keV}$ with a photon index $\Gamma=2$.
The resulting spectral energy distribution peaks above $\sim 1~{\rm keV}$
for BHs below $10^5~\Msol$.
The X-rays photoheat the IGM as they are absorbed.
Because they have long ($\ga 1~{\rm Gpc}$)
mean free paths, they do so nearly isotropically---the feedback is global.\footnote{
Lyman-Werner background radiation from star formation can also
heat the IGM, but \cite{tph12} found this contribution to be subdominant to
the heating due to X-rays from mini-quasars.
}
We solve for the history of this `global warming' of the IGM by miniquasars,
accounting for the atomic ionization states of ${\rm H}$ and ${\rm He}$.
We refer the reader to to \citetalias{tph12} \citep[cf.][]{HaardtMadau96} for the relevant details of cosmological radiative transfer.

As the IGM is heated, its Jeans mass scale increases,
so that gas in low-mass haloes can no longer collapse.
We implement this negative feedback by assuming that
the BHs can only form and grow if their host halo exceeds the Jeans mass,
and that they can only grow for $3\times 10^7~{\rm yr}$ [comparable to
the typical active galactic nucleus (AGN) lifetime; e.g. \citealt{Martini04}]
following the most recent merger of their host with another halo exceeding $M_{\rm J}$.
In this model, the X-ray heating of the IGM by miniquasars is a self-regulatory feedback,
in that the emission of the accreting BHs act to suppress
the subsequent formation and growth of BHs in low-mass haloes.
In models without this self-regulatory feedback,
miniquasars heat the IGM but the IGM temperature has no
bearing on BH formation and growth;
seed BHs form in pristine haloes above the cooling mass
and are allowed to grow continuously at $2/3$ of the Eddington rate.

Because the gas accreted by the BHs following each merger episode
is much less than the fraction of cool gas in the merged halo,
we assume that BSMs do not affect BH growth in our simulations
beyond raising the halo mass threshold above which mergers can trigger accretion.
We discuss caveats to this assumption in \S\ref{s:concl}.

\section{Results}
\label{s:results}

\subsection{Evolution of the cosmic BH population}
BSMs and the negative feedback from IGM heating have the same qualitative effect
of reducing the number of BH seeds,
but they operate preferentially at different epochs.
The former has a greater impact at early times when the relative speeds
between baryons and DM are greater,
and the latter operates at later times after the X-rays from
the earliest growing BHs have sufficiently raised the cosmological Jeans mass. 
To isolate these two effects, we run four sets of simulations,
toggling each effect on and off.
We use the same DM haloes and merger histories for each set of simulations.
The models without regulation by IGM heating vastly overproduce massive BHs
compared to the observational constraints; these are shown
merely to illustrate, in the simplest model of a steadily exponentially growing BHs,
how the late-time BH population is affected when seed BH production is delayed by BSMs.

\begin{figure}
\epsfig{file=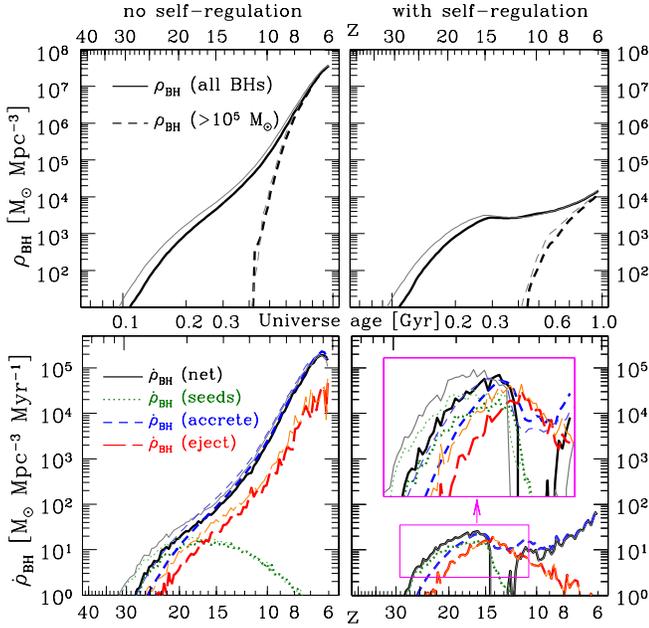,scale=0.44}
\vspace{-0.2in}
\caption{
Evolution of $\rho_{\rm BH}$, the universal mass density of nuclear BHs.
Simulations without BH self-regulation due to IGM heating are presented on the left-hand panels, and
those with self-regulation are presented on the right-hand panels. The thick, darker lines indicate simulations with BSMs included,
and the thin, lighter-coloured lines show those without streaming.
Upper panels: the global mass density of BHs $\rho_{\rm BH}$ as a function of redshift (solid lines),
and the mass density of only the BHs with masses above $10^5~\Msol$ (dashed lines).
Lower panels: the growth rate of the global nuclear BH mass density $\dot{\rho}_{\rm BH}$: net growth (solid black/grey);
growth due to new seed formation (dotted green) and gas accretion (dashed blue)
and losses due to gravitational recoil (long-dashed red).
}
\label{f:rhoBHall}
\end{figure}

We plot the mass evolution of the nuclear BH population in Fig. \ref{f:rhoBHall}.
In all simulations, X-rays from miniquasars `globally warm' the IGM.
The panels on the left-hand side show results from simulations where the rising IGM temperature
provides no feedback on the BHs, whereas the panels on the right-hand side
show results where the warming provides a negative feedback
on the formation and growth of nuclear BHs, as described in the previous section.
The results from simulations implementing BSMs are shown in thick lines,
and those from simulations without BSMs are shown in thin lines with lighter colours.
The solid curves in the upper panels show
the global BH density $\rho_{\rm BH}$ for all nuclear BHs as a function of redshift;
the dashed curves show the density for only the BHs exceeding $10^5~\Msol$.
The lower panels show the rate of change in the universal BH
density $\dot{\rho}_{\rm BH}$ (solid black or grey curves),
along with contributions from new seed formation (dotted green)
and gas accretion (dashed blue),
as well as losses due to gravitational recoil (long-dashed red).
The losses due to BHs being `stuck' in unmerged satellite haloes are not shown.
For the simulations with the `global warming' feedback, the $\dot{\rho}_{\rm BH}$
curves become very crowded near where the IGM heating begins to suppress
seed BH formation and accretion.
We have included a zoom-in view of this region of the plot (magenta boxes, lower right-hand panel)
for ease of viewing.

As anticipated, the primary effect of BSMs is to suppress the formation of seed BHs (PopIII stars)
at early times.
The increase in the effective Jeans mass delays seed BH formation by $\Delta z \sim 3-4$
at $z>20$ \citep{Greif+11a}  in simulations both with and without self-regulation.
A somewhat counterintuitive result is that the suppression of $\rho_{\rm BH}$
by BSMs (relative to the control simulations with BSMs turned off)
decreases with time, even in the simple `no self-regulation' models
in which all BHs grow steadily at $2/3$ of the Eddington rate.
By $z\approx 8$, the total SMBH mass density and its growth rate
in the cases with and without BSMs are nearly indistinguishable.
This is because the `extra' seeds that form in the absence 
of BSMs do not contribute efficiently to $\rho_{\rm BH}$.
Many become satellites as their host haloes become tidally stripped
during a minor merger, and those that do merge with other BHs are often
ejected from the host halo by the gravitational recoil effect.
The vulnerability of the earliest seed BHs to the recoil effect was pointed out by \cite{Haiman04};
\citetalias{TH09} and \citetalias{tph12} had also shown that the number
and masses of massive BHs in merger-tree simulations
depended weakly on the number of seed BHs formed.
In other words, simply doubling (or halving) the number of seed BHs
does not double (or halve) the total mass of nuclear BHs at later times.

\begin{figure}
\epsfig{file=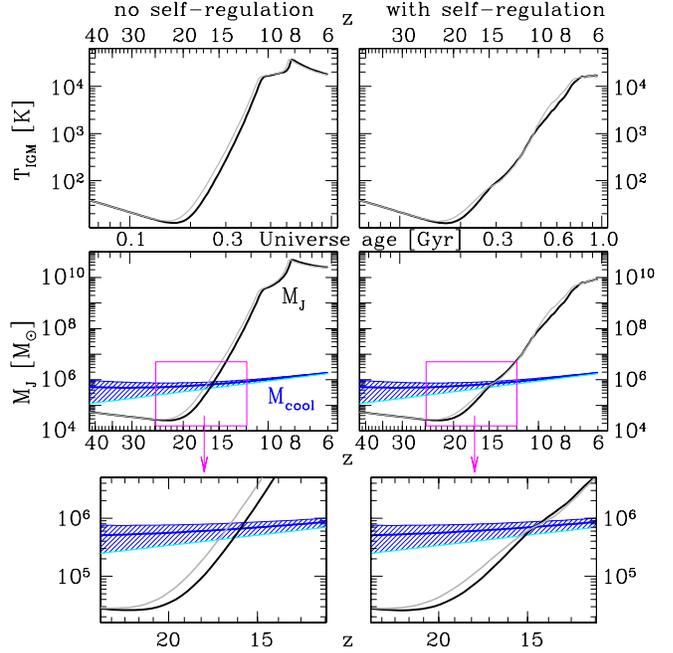,scale=0.44}
\vspace{-0.2in}
\caption{
The thermal history of the IGM in the four simulations.
As with the previous figure, left-hand (right-hand) panels and light (dark) curves
show simulations without (with) self-regulation and without (with) BSMs, respectively.
Top:
the IGM temperature history $T_{\rm IGM}(z)$. The temperature merely rises due to miniquasar
photoheating in the simulations without self-regulation, but in those with self-regulation the temperature
in turn affects the mass scale on which BHs can form and accrete.
Middle and bottom:
the evolution of the cosmological Jeans mass $M_{\rm J}$ as the IGM is heated.
We have shown the distribution (shaded) and mean (solid thick line) of the cooling mass $M_{\rm cool}$
threshold for seed BH formation in blue.
The lower bound of the shaded region, highlighted with a lighter shade of blue,
shows the case without streaming. The upper bound represents a $1\sigma$ deviation from the mean.
The bottom panels show a zoom-in of the region marked by magenta boxes in the middle panels,
just before and after $M_{\rm J}$ overtakes $M_{\rm cool}$.
}
\label{f:IGM}
\end{figure}

The BH mass densities in simulations with and without BSMs
converge earlier and more strongly in the presence of
the global warming, self-regulating feedback by miniquasars.
To illustrate why this occurs, we have plotted in Fig. \ref{f:IGM} the temperature
history of the IGM, $T_{\rm IGM}(z)$ (top panels) in our simulations,
along with the mass scales that govern seed formation and accretion,
$M_{\rm cool}$ and $M_{\rm J}$ (bottom panels).
As with the previous figure, thick black curves show results from the simulations with BSMs,
and thin grey curves show those without BSMs.
We show the distribution of the cooling mass $M_{\rm cool}$ in blue.
The case with no BSMs is highlighted in light blue,
the region between zero streaming velocity and the $1\sigma$
upper bound is shaded in blue, and the mean value of $M_{\rm cool}$
is shown a solid blue curve.
As with Fig. \ref{f:rhoBHall}, we have zoomed in on the region of the figure
where the IGM heating feedback begins to affect BH formation and accretion (bottom panels).

When $M_{\rm J}$ exceeds $M_{\rm cool}$,
the formation of new BHs is rapidly suppressed.
In addition, because the host haloes must exceed the Jeans mass
and have recently merged with another such halo for BHs to
continue growing, the Jeans mass threshold also acts
as a thermostat for BH accretion.
In the simulations without BSMs,
the IGM heats slightly faster than in the case with BSMs,
because there are more BHs to photoheat the IGM.
However, this also means that the Jeans mass rises
faster, and the negative feedback becomes effective earlier.
Once the cosmological Jeans mass exceeds
the cooling mass threshold, BH growth becomes strongly
coupled to the IGM temperature, and BSMs effectively become irrelevant
for BH growth and formation, at least in the context of the models considered here.
This occurs at $z\approx 16$ without BSMs, and $z\approx 14$ with BSMs,
but at $z\le 14$ the IGM temperature (and reionization) histories are virtually indistinguishable.

\subsection{Impact of BSMs on the most massive $z\approx 6$ quasar SMBHs}

\begin{figure}
\epsfig{file=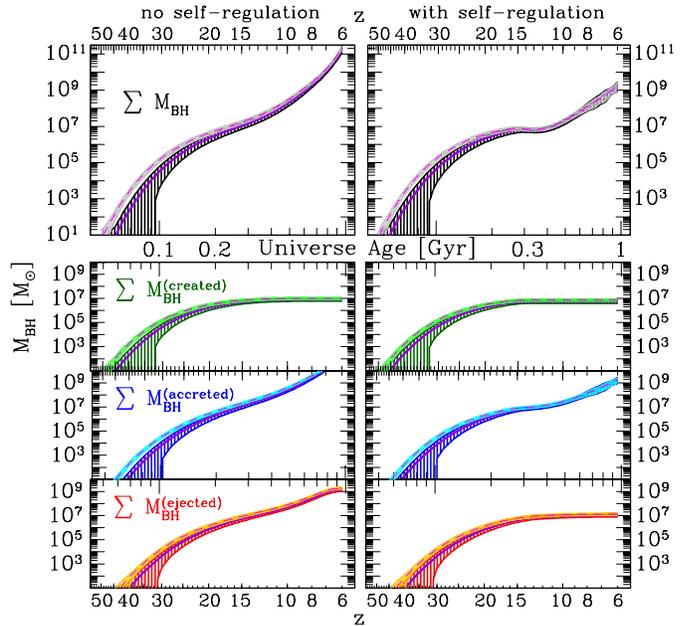,scale=0.44}
\vspace{-0.2in}
\caption{
The distributions of total BH mass per parent halo, as a function of $z$,
for the 70 most massive $z=6$ parent haloes in our simulations.
The $1\sigma$ bounds are shaded, again with
left-hand (right-hand) panels and light (dark) curves
showing simulations without (with) self-regulation and without (with) BSMs, respectively.
The mean of the distributions are shown with purple lines.
BSMs introduce a wide scatter in BH occupation in different sets of progenitors at high $z$,
but at lower redshifts the scatter decreases as the streaming motions decay
and seeds are merged inefficiently due to BHs being stuck as satellites or being ejected via gravitational recoil.
On average, BSMs do not strongly affect the total BH mass in progenitor haloes at $z<15$.
}
\label{f:bigbhs}
\end{figure}

Let us now focus on how BSMs affect the population of the
most massive SMBHs at $z\approx 6$.
We consider the total mass $\sum M_{\rm BH}$
contained in the progenitor haloes that eventually assemble 
a single massive halo at $z=6$ (i.e. the total mass of BHs
inside a given merger tree).
In Fig. \ref{f:bigbhs}, we plot the distribution of this quantity 
for the 70 most massive haloes ($M_{\rm halo}>10^{12.85}~\Msol$)
at $z=6$.
That is, we take the 70 sets of progenitors
(with each set associated with its own streaming velocity value)
that eventually assembly our 70 most massive haloes at $z\approx 6$,
compute $\sum M_{\rm BH} (z)$ for each set, and summarize the distribution.
Here, $\sum M_{\rm BH} (z)$ denotes the total BH mass per parent halo,
not the total BH mass in the $150~{\rm Gpc}^3$ comoving volume of our simulations.
As with the earlier figures, the darker (lighter) shades show the cases with (without) BSMs,
while the left-hand (right-hand) panels show the cases without (with) self-regulation due to IGM heating.
The purple lines in each panel show the mean values per progenitor set,
and these are enveloped by shaded regions that denote the $\pm 1\sigma$ distribution bounds.
The top panels show the total accumulated $\sum M_{\rm BH}$ per parent halo,
and the panels below show the distribution of cumulative BH masses
created as new seeds ($\pm 1\sigma$ bounds shaded in green), accreted (blue)
and ejected via recoil (red).
Because the masses of the parent haloes at $z=6$ differ by less than a factor of 3,
we have opted not to weigh $\sum M_{\rm BH}$ by the parent halo mass.

Note that negative feedback from IGM heating
does not strongly affect the total seed mass created in these haloes (green curves).
This is because most of the seeds are formed before the rise in $T_{\rm IGM}$ and $M_{\rm J}$
turns off seed production. As pointed out by \citetalias{tph12},
these early-forming BHs cause the `global warming' but are largely unaffected by it.

The distribution of $\sum M_{\rm BH}$ has a much narrower spread in the models without streaming velocities.
Note that the thin curves are enveloped by narrow shaded bands denoting the $\pm 1\sigma$ bounds,
whereas the shaded bands around the thick curves are wider (in some cases,
the $1-\sigma$ scatter is larger than the mean, resulting in the shaded regions extending to zero).
Because each of the 70 haloes have millions of seed-forming progenitors,
the total BH formation rate averages out to uniform values in each of these haloes.
The subsequent growth of the BHs is uniform in the models without self-regulation,
and determined by $T_{\rm IGM}(z)$ (which is a global quantity)
and the halo merger history (which averages out over large numbers) in the models with self-regulating feedback.
In the models with BSMs, there is a larger scatter in the seed formation rates
at early times due to statistical variations in $M_{\rm cool}$ from halo to halo,
which then propagates to the subsequent accreted and ejected masses.
However, at late times, the statistical fluctuations decrease
as (i) the spread in $M_{\rm cool}$ decays with the streaming velocities,
and (ii) in the models with self-regulation, the Jeans mass overtakes $M_{\rm cool}$
as the relevant mass scale once the IGM is heated to $\sim 100~{\rm K}$.

We see no evidence in our simulations
that the BSMs impede the `survival of the fittest'
mode of merger-driven growth (\citealt{VolRees06}; \citetalias{TH09}) discussed in \S\ref{sec:intro}.
Naively, one might expect that suppressing the earliest seed BHs
(which would get a head start in growth) would increase the number
of ejected BHs (because mergers between BHs of similar masses
result in larger recoil velocities). 
The number of BH mergers is slightly reduced in the simulations with BSMs
(see also \S\ref{subsec:mergers} and Fig.\ref{f:gw}),
but the fraction of mergers that result in ejection are not appreciably greater.

Fig. \ref{f:bigbhs} reemphasises the qualitative findings discussed earlier:
BSMs can reduce the number of BHs at early times, but this relative suppression
is smeared out at late times because (i) total BH masses have a sublinear
dependence on the total number of seeds formed, since many seeds
are lost as satellites or via gravitational recoil, and (ii) negative feedback due to IGM heating
further tends to suppress the differences at earlier times
by acting as a thermostat for BH formation and growth.
The figure also confirms that while BSMs can be a powerful
suppressant at early times in places where the streaming velocities happen to be large,
the mean effect at late times and across large volumes is much more moderate.

\subsection{Impact of BSMs on the $6<z<11$ quasar luminosity function}

\begin{figure}
\epsfig{file=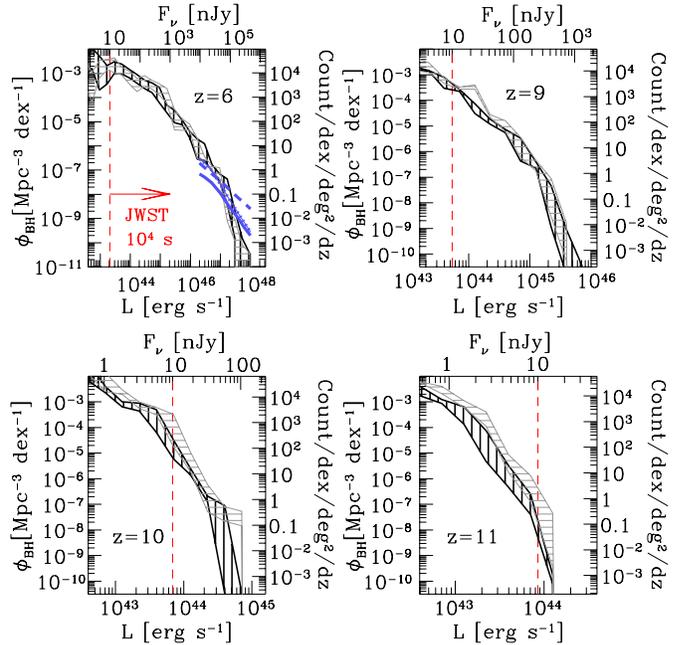,scale=0.44}
\vspace{-0.2in}
\caption{
Quasar luminosity functions in the simulations with self-regulating feedback due to IGM heating,
with (thick, black curves) and without (thin, grey curves) BSMs.
We have selected snapshots of the luminosity function at $z=6$, 9, 10 and 11.
The $z=6$ luminosity functions of  Willott et al. (2010), Shankar et al. (2009),  and Hopkins et al. (2007)
are shown in the upper left-hand panel in solid, dotted and dashed bue curves, respectively.
BSMs appear to suppress the masses/numbers of SMBHs by a factor of few in the range
of $\sim10^6$ to $\sim10^9~\Msol$ ($\sim10^{44}$ to $\sim10^{47}~\erg~\s^{-1}$).
The most drastic difference in the cases with and without BSMs occur
at the most luminous (massive) end of the luminosity function at $z\ga 10$.
While this could reduce the number of luminous quasars that could be detected by \textit{JWST}
at these redshifts, these objects are so rare ($\ltsim 10 ~\deg^{-2}$ per unit redshift at $z=11$)
that they are difficult to find in a blind search in any case.
}
\label{f:qlf}
\end{figure}

Our results so far indicate that BSMs are unlikely to have an appreciable
effect on the SMBH population at $z\sim 6$, at least in the models we have considered.
What about higher redshifts?
We show in Fig. \ref{f:qlf} the quasar luminosity function predicted by the models
with self-regulating global warming feedback, at $z=6$, 9, 10 and 11.
These are shown as shaded regions, to account for the Poisson errors of our simulated sample
and the ambiguity of the duty cycle in our model
(i.e. whether our Eddington ratio of $2/3$
for accreting BHs means that the BHs are shining all the time at $(2/3)L_{\rm Edd}$,
or only $2/3$ of the time at $L_{\rm Edd}$).
As with previous figures, the black, thick curves show the results from simulations with BSMs,
and the grey, thin curves show the ones without.
For convenience, we have also converted the number densities into
the density per square degree per unit redshift, and expressed
the results in terms of the flux density $F_{\nu}$:
\begin{equation}
F_{\nu} = \frac{L}{4\pi d_{\rm L}^2}\frac{1}{b~\nu},
\end{equation}
where $d_L$ is the luminosity distance and $b$ is the bolometric correction.
We have marked the design flux density limit of \textit{JWST}'s
NIRCam instrument ($10^4~{\rm s}$ exposure) at $\approx 3\,\mu{\rm m}$ 
(from \url{http://www.stsci.edu/jwst/overview/design/})
by red vertical lines.
The bolometric correction $b$ can vary from AGN to AGN,
as well as with the intrinsic wavelength;
for example,  \cite{Runnoe+12} report $b\approx 4.2$, $5.2$ and $8.1$
at wavelengths of $1450 \,{\rm \AA}$ (redshifted to $ 3\,\mu{\rm m}$ at $z=19.7$)
$3000 \,{\rm \AA}$ ($z=9$) and $5000 \,{\rm \AA}$ ($z=5$), respectively.
Previous works (see table 2 in \citeauthor{Runnoe+12})
have arrived at values of $b$ within $\sim 30$ per cent of those quoted above.
Given that the typical value of $b$ is unlikely to vary by
more than a factor of two in the redshift range of interest here
($z\approx6$ to $11$, or intrinsic wavelengths of $2500$--$4300\,{\rm \AA}$),
we adopt the constant value $b=5$ for simplicity.

Our results at $z\approx 6$ agree well with the model
luminosity functions of \cite{Hopkins+07b} and \cite{Shankar+09a},
as well as the observationally inferred luminosity function of \cite{Willott+10b},
at $L>10^{46}~\erg~\s^{-1}$, where the data are most robust.
We have plotted these luminosity functions (see fig. 7 of  \citealt{Willott+10b})
in the upper left-hand panel of Fig. \ref{f:qlf} as in blue curves
(\citealt{Willott+10b} in solid, \citealt{Shankar+09a} in dotted, and  \citealt{Hopkins+07b} in dashed curves).
Our models appear to overproduce the less luminous quasars,
but the number density of these dim objects is less certain \citep{Willott+10b}.

For this specific SMBH growth model, BSMs have the largest effect at the high end of the
(mini-) quasar luminosity function out to $z\approx 10$.
Coincidentally, this is comparable to the largest distance at which \textit{JWST}
could detect growing BHs.
It is plausible that models that predict the quasar population at such high redshifts
can overpredict their abundance if they do not take BSMs into account.
However, the suppression affects only the very massive end of the mass function
at $z\ga 10$, i.e. objects whose number densities are $< 1-100 ~{\rm deg}^{-2}$
per unit redshift and thus are extremely difficult to detect even if their
numbers were not suppressed by BSMs.
In our models, BSMs reduce the masses of the most massive miniquasar
BHs by a factor of a few at $z>10$.
At $z\approx 6$, the luminosity functions for the simulations with and without
BSMs are almost indistinguishable.
Our results suggest that the most massive BHs may actually benefit slightly
from BSMs, as the negative effects due to fewer
merging progenitors are outweighed by the positive effect
of weaker X-ray heating at early times.
In any event, the effects of BSMs on the BH mass function at $z<10$
are sufficiently small as to be dwarfed by other theoretical uncertainties.

\subsection{Impact of BSMs on BH merger rates at high $z$}
\label{subsec:mergers}

\begin{figure}
\epsfig{file=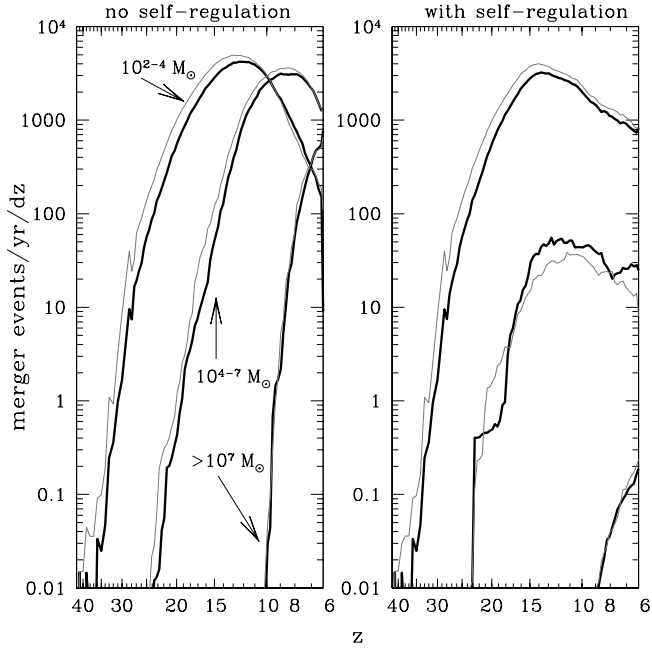,scale=0.44}
\vspace{-0.2in}
\caption{
The merger rate of nuclear BHs with mass ratios $M_2/M_1>0.01$.
As with Figures \ref{f:rhoBHall} through \ref{f:bigbhs},
left-hand (right-hand) panels and light (dark) curves
show results from simulations without (with) self-regulating feedback and BSMs, respectively.
In each panel, three total mass bins are shown: $10^2<(1+z)M/\Msol<10^4$ (top/left curves),
 $10^4<(1+z)M/\Msol<10^7$ (middle curves, corresponding roughly to the \textit{eLISA} window)
 and $(1+z)M/\Msol>10^7$ (bottom/right curves).
 BSMs are unlikely to affect estimates for the detection rates
 for proposed space-based gravitational-wave detectors
 (see caveats in \S\ref{s:concl}).}
\label{f:gw}
\end{figure}

Finally, we address whether BSMs could impact the number of mergers of massive BHs
in the early Universe. Such mergers are primary targets
of proposed space-based gravitational-wave detectors such as
\textit{eLISA}\footnote{\url{https://www.elisascience.org}} \citep{eLISApaper}
 and \textit{DECIGO} \citep{DECIGOpaper}.
In Fig. \ref{f:gw} we show the merger rates of BHs in the redshifted binary mass ranges\footnote{
The factor $1+z$ arises from the degeneracy between the mass
and redshift of a gravitational wave source \citep[e.g.][]{Hughes02}.}
$10^{2}~\Msol<(1+z)M<10^4~\Msol$ (top/left curves),
$10^{4}~\Msol<(1+z)M<10^7~\Msol$ (middle curves, approximately
coincident with the \textit{eLISA} sensitivity window), and
$(1+z)M>10^7~\Msol$ (bottom/right curves).
We have considered binaries with a minimum mass ratio $M_2/M_1>0.01$,
neglecting BHs in satellite haloes whose dynamical merger times are longer
than the Hubble time.
As with Figures \ref{f:rhoBHall} through \ref{f:bigbhs}, left-hand (right-hand)
panels and thin grey (thick black)
curves denote models without (with) self-regulating feedback from IGM heatingand
streaming velocities, respectively.
For the models considered, BSMs do not affect the expected merger
rate by more than a factor of 2 at $z<15$.
Once again, our results suggest that BSMs will not appreciably
affect the observability of nuclear BHs in the early Universe.

\section{Conclusions}
\label{s:concl}

We have investigated the possible consequences of the early suppression
of the formation of seed BHs by the relative bulk motion of baryons
against DM haloes in the early Universe.
We investigated how BSMs can affect the population of SMBHs at later
times (out to $z=6$) by way of a toy model in which all
BHs grow at $2/3$ of the Eddington limit, and a more realistic one in which they
do so only for a short period after their host haloes experienced a merger
in which both haloes exceed the cosmological Jeans mass.
In both pairs of models, we found that the effects of BSMs
on observable quantities---(i) the total mass density of nuclear BHs, (ii) their mass function,
(iii) their contributions to photoheating the IGM,
and (iv) their merger rates---to be marginal. The only exception
is that we found the BSMs could suppress the masses of the most massive BHs
at $z>10$; however, the suppression occurs only for 
the rarest objects that are unlikely to be found by \textit{JWST} in any case.
Our models suggest that the suppression of the nuclear BH population
could be much greater at $z\ga 20$, but it is unclear
whether this has observational consequences.
Although we have only considered PopIII seeds,
a simple extrapolation of our findings suggests that models with more massive seeds,
in which seeds form later (e.g. \citealt{Begelman+06}, \citealt{LodatoNatarajan06},
\citealt{Latif+13}, \citealt{Schleicher+13}),
would be even less affected by BSMs.

An important caveat to the above summary is that
we have assumed that the suppression by BSMs of the cold gas content
in DM minihaloes only affects BH seed formation, and not their subsequent growth.
Essentially, we have assumed that the amount of gas in the host halo is not a bottleneck for BH growth.
There are extreme examples, such as present-day dwarf galaxies
or the $z\sim 10$ minihaloes far below the cosmological Jeans mass
in our simulations, where the severe dearth of cool gas can almost completely
inhibit BH growth; but the suppression by BSMs of the cold gas fraction
and central gas densities at $z\sim 20$
is thought to typically only be a factor of a few (e.g. \citealt{Greif+11a}, \citealt{Naoz+13}).
Since nuclear BHs typically consume less than $1$ per cent of the total baryonic
mass in their host haloes, the factors that limit their growth are
unlikely to be directly related to factors of a few in the
amount of gas in the halo, but rather to opportunity (e.g. frequency of major mergers of the host halo)
or other regulatory mechanisms (local or global feedback).
The nuclear BH population may be more sensitive to suppression by BSMs
if BH masses tend to scale strongly with the cold gas fraction of the host halo,
or if growing BHs is somehow extremely sensitive to the gas content of the host
at $z\ga 15$ (e.g. through density and temperature thresholds for feedback or secular
instabilities that could help fuel BH growth).
Additionally, \cite{Greif+11a} found that BSMs can help drive turbulence
inside minihaloes, which can impact the gas accretion rates
and thus affect both the PopIII star masses as well as the growth of the
seed BHs they leave behind \citep[e.g.][]{Krumholz+06}.

Another caveat is the use of the cosmological Jeans mass in regulating BH
formation and growth. It has been suggested that the more pertinent
mass scale may instead be the so-called filtering mass, which depends on
the prior temperature history of the IGM \citep{Gnedin00}.
The Jeans mass may overestimate by a factor of a few the halo mass scale
on which negative feedback from IGM heating becomes effective \citep[e.g.][]{Naoz+13}

In a similar vein, if the merger time-scales of BHs depend sensitively
on the gas content and density profiles of their host (e.g. \citealt{Mayer+07}),
then BSMs may act to suppress or flatten the
BH merger event rate as a function of redshift.

We have also not included in our merger trees the suppression
of DM halo abundances by BSMs. The results of \cite{fbth12} suggest that this
effect may be as important at $z\ltsim 30$ as the increase in $M_{\rm min}$
(the threshold halo mass for PopIII star formation). Accounting for this
effect would also somewhat increase the impact of the BSMs.
However, considering that this suppression is about a factor of 2 at $z\ltsim 30$
and that its relative importance is much lower at $z\ga 30$ when BSMs are greatest,
it is unlikely  to change our basic conclusion.

We conclude that unless SMBH growth and mergers are extremely sensitive
to order-unity fluctuations in the cold gas density of their host haloes,
BSMs are unlikely to play a dramatic role in suppressing the
formation of the first SMBHs. They may reduce the luminosities (masses)
of the most luminous quasars at $z>10$ by a factor of a few, but otherwise
we find the overall average effect on the nuclear BH population and their observables
at $6\ltsim z \ltsim 10$ to be marginal.
BSMs could, however, still leave detectable imprints on small ($\ltsim 100~{\rm Mpc}$) scales.
If quasars and galaxies are systematically less luminous in regions affected by BSMs,
such spatial correlations could be detectable in future large-area, high-redshift surveys
that measure clustering properties of quasars and galaxies (see \citealt{Dalal+10}, \citealt{tbh11}).
Similarly, spatial correlations in quasar luminosity
and star formation can cause reionization to take place more inhomogeneously,
which could leave detectable imprints in the 21cm power spectrum \citep{Maio+11, McQuinnOleary12, Visbal+12}.

\section*{Acknowledgements}
The authors are grateful to Jun Zhang for helpful discussions regarding the merger-tree algorithm
and to Ryan O'Leary for discussions of the literature.
We thank the anonymous referee, whose careful comments improved the clarity of this work.
This work was partially supported by NASA grant NNX11AE05G (to ZH).


\begin{thebibliography}{76}
\expandafter\ifx\csname natexlab\endcsname\relax\def\natexlab#1{#1}\fi

\bibitem[{{Ade} {et~al.}(2013){Ade}, {Aghanim}, {Armitage-Caplan}, {Arnaud},
  {Ashdown}, {Atrio-Barandela}, {Aumont}, {Baccigalupi}, {Banday}, \&
  et~al.}]{PlanckParameters}
{Ade}, P.~A.~R. {et~al.} 2013, ArXiv
  e-prints 1303.5076

\bibitem[{{Alvarez} {et~al.}(2009){Alvarez}, {Wise}, \& {Abel}}]{Alvarez+09}
{Alvarez}, M.~A., {Wise}, J.~H., \& {Abel}, T. 2009, \apjl, 701, L133

\bibitem[{{Baker} {et~al.}(2006){Baker}, {Centrella}, {Choi}, {Koppitz}, {van
  Meter}, \& {Miller}}]{Baker+06b}
{Baker}, J.~G., {Centrella}, J., {Choi}, D., Koppitz M., van Meter J. R., Miller M. C.,2006, \apjl, 653, L93

\bibitem[{{Barkana} \& {Loeb}(2001)}]{BarkanaLoeb01}
{Barkana}, R., \& {Loeb}, A. 2001, Phys. Rep., 349, 125

\bibitem[{{Begelman} {et~al.}(2006){Begelman}, {Volonteri}, \&
  {Rees}}]{Begelman+06}
{Begelman}, M.~C., {Volonteri}, M., \& {Rees}, M.~J. 2006, \mnras, 370, 289

\bibitem[{{Blaes}(2004)}]{Blaes04}
{Blaes}, O.~M. 2004, in Beskin V., Henri G., Menard F.,
Pelletier G., Dalibard J., eds, Accretion
Discs, Jets and High Energy Phenomena in Astrophysics.
Springer, New York, p. 137-185

\bibitem[{{Bogdanovi{\'c}} {et~al.}(2007){Bogdanovi{\'c}}, {Reynolds}, \&
  {Miller}}]{Bogdanovic+07}
{Bogdanovi{\'c}}, T., {Reynolds}, C.~S., \& {Miller}, M.~C. 2007, \apjl, 661,
  L147

\bibitem[{{Boylan-Kolchin} {et~al.}(2008){Boylan-Kolchin}, {Ma}, \&
  {Quataert}}]{BoylanKolchin+08}
{Boylan-Kolchin}, M., {Ma}, C., \& {Quataert}, E. 2008, \mnras, 383, 93

\bibitem[{{Bromley} {et~al.}(2004){Bromley}, {Somerville}, \&
  {Fabian}}]{Bromley+04}
{Bromley}, J.~M., {Somerville}, R.~S., \& {Fabian}, A.~C. 2004, \mnras, 350,
  456

\bibitem[{{Bromm} \& {Larson}(2004)}]{bl04}
{Bromm}, V., \& {Larson}, R.~B. 2004, ARA\&A, 42, 79

\bibitem[{{Bromm} {et~al.}(2009){Bromm}, {Yoshida}, {Hernquist}, \&
  {McKee}}]{Bromm+09}
{Bromm}, V., {Yoshida}, N., {Hernquist}, L., \& {McKee}, C.~F. 2009, \nat, 459,
  49

\bibitem[{{Carroll} {et~al.}(1992){Carroll}, {Press}, \& {Turner}}]{cpt92}
{Carroll}, S.~M., {Press}, W.~H., \& {Turner}, E.~L. 1992, ARA\&A, 30, 499

\bibitem[{{Dalal} {et~al.}(2010){Dalal}, {Pen}, \& {Seljak}}]{Dalal+10}
{Dalal}, N., {Pen}, U.-L., \& {Seljak}, U. 2010, J. Cosmol. Astropart. Phys., 11, 7

\bibitem[{{Danzmann} {et~al.}(2013){Danzmann}, {Seoane}, {Aoudia}, {Audley},
  {Auger}, {Babak}, {Baker}, {Barausse}, {Barke}, {Bassan}, {Beckmann},
  {Benacquista}, {Bender}, {Berti}, {Bin{\'e}truy}, {Bogenstahl}, {Bonvin},
  {Bortoluzzi}, {Brause}, {Brossard}, {Buchman}, {Bykov}, {Camp}, {Caprini},
  {Cavalleri}, {Cerdonio}, {Ciani}, {Colpi}, {Congedo}, {Conklin}, {Cornish},
  {de Vine}, {DeBra}, {Dewi Freitag}, {Di Fiore}, {Diaz Aguilo}, {Diepholz},
  {Dolesi}, {Dotti}, {Fern{\'a}ndez Barranco}, {Ferraioli}, {Ferroni},
  {Finetti}, {Fitzsimons}, {Gair}, {Galeazzi}, {Garcia}, {Gerberding}, {Gesa},
  {Giardini}, {Gibert}, {Grimani}, {Groot}, {Guzman Cervantes}, {Haiman},
  {Halloin}, {Heinzel}, {Hewitson}, {Hogan}, {Holz}, {Hornstrup}, {Hoyland},
  {Hoyle}, {Hueller}, {Hughes}, {Jetzer}, {Kalogera}, {Karnesis}, {Kilic},
  {Killow}, {Klipstein}, {Kochkina}, {Korsakova}, {Krolak}, {Larson}, {Lieser},
  {Littenberg}, {Livas}, {Lloro}, {Mance}, {Madau}, {Maghami}, {Mahrdt},
  {Marsh}, {Mateos}, {Mayer}, {McClelland}, {McKenzie}, {McWilliams},
  {Merkowitz}, {Miller}, {Mitryk}, {Moerschell}, {Mohanty}, {Monsky},
  {Mueller}, {M{\"u}ller}, {Nelemans}, {Nicolodi}, {Nissanke}, {Nofrarias},
  {Numata}, {Ohme}, {Otto}, {Perreur-Lloyd}, {Petiteau}, {Phinney}, {Plagnol},
  {Pollack}, {Porter}, {Prat}, {Preston}, {Prince}, {Reiche}, {Richstone},
  {Robertson}, {Rossi}, {Rosswog}, {Rubbo}, {Ruiter}, {Sanjuan},
  {Sathyaprakash}, {Schlamminger}, {Schutz}, {Sch{\"u}tze}, {Sesana},
  {Shaddock}, {Shah}, {Sheard}, {Sopuerta}, {Spector}, {Spero}, {Stanga},
  {Stebbins}, {Stede}, {Steier}, {Sumner}, {Sun}, {Sutton}, {Tanaka}, {Tanner},
  {Thorpe}, {Tr{\"o}bs}, {Tinto}, {Tu}, {Vallisneri}, {Vetrugno}, {Vitale},
  {Volonteri}, {Wand}, {Wang}, {Wanner}, {Ward}, {Ware}, {Wass}, {Weber}, {Yu},
  {Yunes}, \& {Zweifel}}]{eLISApaper}
{Danzmann}, K. {et~al.} 2013, ArXiv e-prints 1305.5720

\bibitem[{{Dotti} {et~al.}(2009){Dotti}, {Montuori}, {Decarli}, {Volonteri},
  {Colpi}, \& {Haardt}}]{Dotti+09}
{Dotti}, M., {Montuori}, C., {Decarli}, R., Volonteri M., Colpi M., Haardt F., 2009, \mnras, 398, L73

\bibitem[{{Eisenstein} \& {Hu}(1998)}]{eh98}
{Eisenstein}, D.~J., \& {Hu}, W. 1998, \apj, 496, 605

\bibitem[{{Fan} {et~al.}(2001){Fan}, {Narayanan}, {Lupton}, {Strauss}, {Knapp},
  {Becker}, {White}, {Pentericci}, {Leggett}, {Haiman}, {Gunn}, {Ivezi{\'c}},
  {Schneider}, {Anderson}, {Brinkmann}, {Bahcall}, {Connolly}, {Csabai}, {Doi},
  {Fukugita}, {Geballe}, {Grebel}, {Harbeck}, {Hennessy}, {Lamb}, {Miknaitis},
  {Munn}, {Nichol}, {Okamura}, {Pier}, {Prada}, {Richards}, {Szalay}, \&
  {York}}]{fan01}
{Fan}, X. {et~al.} 2001, \aj, 122, 2833

\bibitem[{{Fan} {et~al.}(2006){Fan}, {Strauss}, {Becker}, {White}, {Gunn},
  {Knapp}, {Richards}, {Schneider}, {Brinkmann}, \& {Fukugita}}]{fan06}
{Fan}, X. {et~al.} 2006, \aj, 132, 117

\bibitem[{{Favata} {et~al.}(2004){Favata}, {Hughes}, \& {Holz}}]{Favata+04}
{Favata}, M., {Hughes}, S.~A., \& {Holz}, D.~E. 2004, \apjl, 607, L5

\bibitem[{{Fialkov} {et~al.}(2012){Fialkov}, {Barkana}, {Tseliakhovich}, \&
  {Hirata}}]{fbth12}
{Fialkov}, A., {Barkana}, R., {Tseliakhovich}, D., \& {Hirata}, C.~M. 2012,
  \mnras, 424, 1335

\bibitem[{{Gnedin}(2000)}]{Gnedin00}
{Gnedin}, N.~Y. 2000, \apj, 542, 535



\bibitem[{{Greif} {et~al.}(2011{\natexlab{a}}){Greif}, {White}, {Klessen}, \&
  {Springel}}]{Greif+11a}
{Greif}, T.~H., {White}, S.~D.~M., {Klessen}, R.~S., \& {Springel}, V.
  2011{\natexlab{a}}, \apj, 736, 147

\bibitem[{{Greif} {et~al.}(2011{\natexlab{b}}){Greif}, {Springel}, {White},
  {Glover}, {Clark}, {Smith}, {Klessen}, \& {Bromm}}]{Greif+11b}
{Greif}, T.~H., {Springel}, V., {White}, S.~D.~M., Glover S. C. O.,
Clark P. C., Smith R. J., Klessen
R. S., Bromm V., 2011{\natexlab{b}},
  \apj, 737, 75

\bibitem[{{Haardt} \& {Madau}(1996)}]{HaardtMadau96}
{Haardt}, F., \& {Madau}, P. 1996, \apj, 461, 20

\bibitem[{{Haiman}(2004)}]{Haiman04}
{Haiman}, Z. 2004, \apj, 613, 36

\bibitem[{{Haiman}(2013)}]{Haiman13}
{Haiman}, Z. 2013, in Wiklind T., Mobasher B., Bromm V., eds, Astrophysics and Space Science
Library, Vol. 396,The First Galaxies. Springer, Berlin, p.293

\bibitem[{{Haiman} \& {Loeb}(2001)}]{HaimanLoeb01}
{Haiman}, Z., \& {Loeb}, A. 2001, \apj, 552, 459

\bibitem[{{Heger} {et~al.}(2003){Heger}, {Fryer}, {Woosley}, {Langer}, \&
  {Hartmann}}]{Heger+03}
{Heger}, A., {Fryer}, C.~L., {Woosley}, S.~E., {Langer}, N., \& {Hartmann},
  D.~H. 2003, \apj, 591, 288

\bibitem[{{Hinshaw} {et~al.}(2012){Hinshaw}, {Larson}, {Komatsu}, {Spergel},
  {Bennett}, {Dunkley}, {Nolta}, {Halpern}, {Hill}, {Odegard}, {Page}, {Smith},
  {Weiland}, {Gold}, {Jarosik}, {Kogut}, {Limon}, {Meyer}, {Tucker}, {Wollack},
  \& {Wright}}]{Hinshaw+12}
{Hinshaw}, G. {et~al.} 2012, ArXiv e-prints
  1212.5226

\bibitem[{{Hopkins} {et~al.}(2007){Hopkins}, {Richards}, \&
  {Hernquist}}]{Hopkins+07b}
{Hopkins}, P.~F., {Richards}, G.~T., \& {Hernquist}, L. 2007, \apj, 654, 731

\bibitem[{{Hosokawa} {et~al.}(2011){Hosokawa}, {Omukai}, {Yoshida}, \&
  {Yorke}}]{Hosokawa+11}
{Hosokawa}, T., {Omukai}, K., {Yoshida}, N., \& {Yorke}, H.~W. 2011, Science,
  334, 1250

\bibitem[{{Hughes}(2002)}]{Hughes02}
{Hughes}, S.~A. 2002, \mnras, 331, 805

\bibitem[{{Jarosik} {et~al.}(2011){Jarosik}, {Bennett}, {Dunkley}, {Gold},
  {Greason}, {Halpern}, {Hill}, {Hinshaw}, {Kogut}, {Komatsu}, {Larson},
  {Limon}, {Meyer}, {Nolta}, {Odegard}, {Page}, {Smith}, {Spergel}, {Tucker},
  {Weiland}, {Wollack}, \& {Wright}}]{Jarosik+11}
{Jarosik}, N. {et~al.} 2011, \apjs, 192, 14

\bibitem[{{Kawamura} {et~al.}(2011){Kawamura}, {Ando}, {Seto}, {Sato},
  {Nakamura}, {Tsubono}, {Kanda}, {Tanaka}, {Yokoyama}, {Funaki}, {Numata},
  {Ioka}, {Takashima}, {Agatsuma}, {Akutsu}, {Aoyanagi}, {Arai}, {Araya},
  {Asada}, {Aso}, {Chen}, {Chiba}, {Ebisuzaki}, {Ejiri}, {Enoki}, {Eriguchi},
  {Fujimoto}, {Fujita}, {Fukushima}, {Futamase}, {Harada}, {Hashimoto},
  {Hayama}, {Hikida}, {Himemoto}, {Hirabayashi}, {Hiramatsu}, {Hong},
  {Horisawa}, {Hosokawa}, {Ichiki}, {Ikegami}, {Inoue}, {Ishidoshiro},
  {Ishihara}, {Ishikawa}, {Ishizaki}, {Ito}, {Itoh}, {Izumi}, {Kawano},
  {Kawashima}, {Kawazoe}, {Kishimoto}, {Kiuchi}, {Kobayashi}, {Kohri},
  {Koizumi}, {Kojima}, {Kokeyama}, {Kokuyama}, {Kotake}, {Kozai}, {Kunimori},
  {Kuninaka}, {Kuroda}, {Kuroyanagi}, {Maeda}, {Matsuhara}, {Matsumoto},
  {Michimura}, {Miyakawa}, {Miyamoto}, {Miyoki}, {Morimoto}, {Morisawa},
  {Moriwaki}, {Mukohyama}, {Musha}, {Nagano}, {Naito}, {Nakamura}, {Nakano},
  {Nakao}, {Nakasuka}, {Nakayama}, {Nakazawa}, {Nishida}, {Nishiyama},
  {Nishizawa}, {Niwa}, {Noumi}, {Obuchi}, {Ohashi}, {Ohishi}, {Ohkawa},
  {Okada}, {Okada}, {Oohara}, {Sago}, {Saijo}, {Saito}, {Sakagami}, {Sakai},
  {Sakata}, {Sasaki}, {Sato}, {Shibata}, {Shinkai}, {Shoda}, {Somiya},
  {Sotani}, {Sugiyama}, {Suwa}, {Suzuki}, {Tagoshi}, {Takahashi}, {Takahashi},
  {Takahashi}, {Takahashi}, {Takahashi}, {Takahashi}, {Takahashi}, {Akiteru},
  {Takano}, {Tanaka}, {Taniguchi}, {Taruya}, {Tashiro}, {Torii}, {Toyoshima},
  {Tsujikawa}, {Tsunesada}, {Ueda}, {Ueda}, {Utashima}, {Wakabayashi}, {Yagi},
  {Yamakawa}, {Yamamoto}, {Yamazaki}, {Yoo}, {Yoshida}, {Yoshino}, \&
  {Sun}}]{DECIGOpaper}
{Kawamura}, S., {Ando}, M., {Seto}, N., {et~al.} 2011, Class. Quantum
  Gravity, 28, 094011

\bibitem[{{Kidder}(1995)}]{Kidder95}
{Kidder}, L.~E. 1995, \prd, 52, 821

\bibitem[{{Krumholz} {et~al.}(2006){Krumholz}, {McKee}, \&
  {Klein}}]{Krumholz+06}
{Krumholz}, M.~R., {McKee}, C.~F., \& {Klein}, R.~I. 2006, \apj, 638, 369

\bibitem[{{Lacey} \& {Cole}(1993)}]{LaceyCole93}
{Lacey}, C., \& {Cole}, S. 1993, \mnras, 262, 627

\bibitem[{{Latif} {et~al.}(2013){Latif}, {Schleicher}, {Schmidt}, \&
  {Niemeyer}}]{Latif+13}
{Latif}, M.~A., {Schleicher}, D.~R.~G., {Schmidt}, W., \& {Niemeyer}, J. 2013,
  \mnras, 433, 1607

\bibitem[{{Lawrence} {et~al.}(2007){Lawrence}, {Warren}, {Almaini}, {Edge},
  {Hambly}, {Jameson}, {Lucas}, {Casali}, {Adamson}, {Dye}, {Emerson},
  {Foucaud}, {Hewett}, {Hirst}, {Hodgkin}, {Irwin}, {Lodieu}, {McMahon},
  {Simpson}, {Smail}, {Mortlock}, \& {Folger}}]{lawrence07}
{Lawrence}, A., {Warren}, S.~J., {Almaini}, O., {et~al.} 2007, \mnras, 379,
  1599

\bibitem[{{Lodato} \& {Natarajan}(2006)}]{LodatoNatarajan06}
{Lodato}, G., \& {Natarajan}, P. 2006, \mnras, 371, 1813

\bibitem[{{Lousto} {et~al.}(2010){Lousto}, {Campanelli}, {Zlochower}, \&
  {Nakano}}]{lousto10}
{Lousto}, C.~O., {Campanelli}, M., {Zlochower}, Y., \& {Nakano}, H. 2010,
  Class. Quantum Gravity, 27, 114006

\bibitem[{{Madau} \& {Rees}(2001)}]{MadauRees01}
{Madau}, P., \& {Rees}, M.~J. 2001, \apjl, 551, L27

\bibitem[{{Maio} {et~al.}(2011){Maio}, {Koopmans}, \& {Ciardi}}]{Maio+11}
{Maio}, U., {Koopmans}, L.~V.~E., \& {Ciardi}, B. 2011, \mnras, 412, L40

\bibitem[{{Martini}(2004)}]{Martini04}
{Martini}, P. 2004, in Coevolution of Black Holes and Galaxies. ed. {L.~C.~Ho}
  ({Cambridge Univ. Press, Cambridge}), p.169

\bibitem[{{Mayer} {et~al.}(2007){Mayer}, {Kazantzidis}, {Madau}, {Colpi},
  {Quinn}, \& {Wadsley}}]{Mayer+07}
{Mayer}, L., {Kazantzidis}, S., {Madau}, P., Colpi M., Quinn T., Wadsley J., 2007, Science, 316, 1874

\bibitem[{{McQuinn} \& {O'Leary}(2012)}]{McQuinnOleary12}
{McQuinn}, M., \& {O'Leary}, R.~M. 2012, \apj, 760, 3

\bibitem[{{Milosavljevi{\'c}} {et~al.}(2009){Milosavljevi{\'c}}, {Bromm},
  {Couch}, \& {Oh}}]{Milos+09}
{Milosavljevi{\'c}}, M., {Bromm}, V., {Couch}, S.~M., \& {Oh}, S.~P. 2009,
  \apj, 698, 766

\bibitem[{{Mortlock} {et~al.}(2011){Mortlock}, {Warren}, {Venemans}, {Patel},
  {Hewett}, {McMahon}, {Simpson}, {Theuns}, {Gonz{\'a}les-Solares}, {Adamson},
  {Dye}, {Hambly}, {Hirst}, {Irwin}, {Kuiper}, {Lawrence}, \&
  {R{\"o}ttgering}}]{mortlock11}
{Mortlock}, D.~J. {et~al.} 2011, \nat,
  474, 616

\bibitem[{{Naoz} {et~al.}(2013){Naoz}, {Yoshida}, \& {Gnedin}}]{Naoz+13}
{Naoz}, S., {Yoshida}, N., \& {Gnedin}, N.~Y. 2013, \apj, 763, 27

\bibitem[{{Ohkubo} {et~al.}(2009){Ohkubo}, {Nomoto}, {Umeda}, {Yoshida}, \&
  {Tsuruta}}]{Ohkubo+09}
{Ohkubo}, T., {Nomoto}, K., {Umeda}, H., {Yoshida}, N., \& {Tsuruta}, S. 2009,
  \apj, 706, 1184

\bibitem[{{Omukai} \& {Palla}(2001)}]{OmukaiPalla03}
{Omukai}, K., \& {Palla}, F. 2001, \apjl, 561, L55

\bibitem[{{Pelupessy} {et~al.}(2007){Pelupessy}, {Di Matteo}, \&
  {Ciardi}}]{Pelupessy+07}
{Pelupessy}, F.~I., {Di Matteo}, T., \& {Ciardi}, B. 2007, \apj, 665, 107

\bibitem[{{Peres}(1962)}]{Peres62}
{Peres}, A. 1962, Phys. Rev., 128, 2471

\bibitem[{{Runnoe} {et~al.}(2012){Runnoe}, {Brotherton}, \&
  {Shang}}]{Runnoe+12}
{Runnoe}, J.~C., {Brotherton}, M.~S., \& {Shang}, Z. 2012, \mnras, 422, 478

\bibitem[{{Schleicher} {et~al.}(2013){Schleicher}, {Palla}, {Ferrara}, {Galli},
  \& {Latif}}]{Schleicher+13}
{Schleicher}, D.~R.~G., {Palla}, F., {Ferrara}, A., {Galli}, D., \& {Latif}, M.
  2013, ArXiv e-prints 1305.5923

\bibitem[{{Shakura} \& {Sunyaev}(1973)}]{ss73}
{Shakura}, N.~I., \& {Sunyaev}, R.~A. 1973, \aap, 24, 337

\bibitem[{{Shankar} {et~al.}(2009){Shankar}, {Weinberg}, \&
  {Miralda-Escud{\'e}}}]{Shankar+09a}
{Shankar}, F., {Weinberg}, D.~H., \& {Miralda-Escud{\'e}}, J. 2009, \apj, 690,
  20

\bibitem[{{Shapiro}(2005)}]{Shapiro05}
{Shapiro}, S.~L. 2005, \apj, 620, 59

\bibitem[{{Sheth} \& {Tormen}(2002)}]{st02}
{Sheth}, R.~K., \& {Tormen}, G. 2002, \mnras, 329, 61

\bibitem[{{Stacy} {et~al.}(2011){Stacy}, {Bromm}, \& {Loeb}}]{sbl11}
{Stacy}, A., {Bromm}, V., \& {Loeb}, A. 2011, \apjl, 730, L1

\bibitem[{{Stacy} {et~al.}(2010){Stacy}, {Greif}, \& {Bromm}}]{Stacy+10}
{Stacy}, A., {Greif}, T.~H., \& {Bromm}, V. 2010, \mnras, 403, 45

\bibitem[{{Tanaka} \& {Haiman}(2009)}]{TH09}
{Tanaka}, T., \& {Haiman}, Z. 2009, \apj, 696, 1798 (TH09)

\bibitem[{{Tanaka} \& {Menou}(2010)}]{TM10}
{Tanaka}, T., \& {Menou}, K. 2010, \apj, 714, 404

\bibitem[{{Tanaka} {et~al.}(2012){Tanaka}, {Perna}, \& {Haiman}}]{tph12}
{Tanaka}, T., {Perna}, R., \& {Haiman}, Z. 2012, \mnras, 425, 2974 (TPH12)

\bibitem[{{Tseliakhovich} {et~al.}(2011){Tseliakhovich}, {Barkana}, \&
  {Hirata}}]{tbh11}
{Tseliakhovich}, D., {Barkana}, R., \& {Hirata}, C.~M. 2011, \mnras, 418, 906

\bibitem[{{Tseliakhovich} \& {Hirata}(2010)}]{th10}
{Tseliakhovich}, D., \& {Hirata}, C. 2010, Phys. Rev. D, 82, 083520

\bibitem[{{Turk} {et~al.}(2009){Turk}, {Abel}, \& {O'Shea}}]{Turk+09}
{Turk}, M.~J., {Abel}, T., \& {O'Shea}, B. 2009, Science, 325, 601

\bibitem[{{Visbal} {et~al.}(2012){Visbal}, {Barkana}, {Fialkov},
  {Tseliakhovich}, \& {Hirata}}]{Visbal+12}
{Visbal}, E., {Barkana}, R., {Fialkov}, A., {Tseliakhovich}, D., \& {Hirata},
  C.~M. 2012, \nat, 487, 70

\bibitem[{{Volonteri}(2010)}]{volonteri10}
{Volonteri}, M. 2010, \aapr, 18, 279

\bibitem[{{Volonteri} {et~al.}(2003){Volonteri}, {Haardt}, \& {Madau}}]{VHM03}
{Volonteri}, M., {Haardt}, F., \& {Madau}, P. 2003, \apj, 582, 559

\bibitem[{{Volonteri} \& {Rees}(2006)}]{VolRees06}
{Volonteri}, M., \& {Rees}, M.~J. 2006, \apj, 650, 669

\bibitem[{{Willott} {et~al.}(2009){Willott}, {Delorme}, {Reyl{\'e}}, {Albert},
  {Bergeron}, {Crampton}, {Delfosse}, {Forveille}, {Hutchings}, {McLure},
  {Omont}, \& {Schade}}]{willott09}
{Willott}, C.~J. {et~al.} 2009, \aj, 137, 3541

\bibitem[{{Willott} {et~al.}(2010){Willott}, {Albert}, {Arzoumanian},
  {Bergeron}, {Crampton}, {Delorme}, {Hutchings}, {Omont}, {Reyl{\'e}}, \&
  {Schade}}]{Willott+10b}
{Willott}, C.~J. {et~al.} 2010, \aj, 140, 546

\bibitem[{{Yoo} \& {Miralda-Escud{\'e}}(2004)}]{YooME04}
{Yoo}, J., \& {Miralda-Escud{\'e}}, J. 2004, \apjl, 614, L25

\bibitem[{{Zhang} {et~al.}(2008{\natexlab{a}}){Zhang}, {Ma}, \&
  {Fakhouri}}]{zmf08}
{Zhang}, J., {Ma}, C.-P., \& {Fakhouri}, O. 2008{\natexlab{a}}, \mnras, 387,
  L13

\bibitem[{{Zhang} {et~al.}(2008{\natexlab{b}}){Zhang}, {Woosley}, \&
  {Heger}}]{ZhangW+08}
{Zhang}, W., {Woosley}, S.~E., \& {Heger}, A. 2008{\natexlab{b}}, \apj, 679,
  639

\end{thebibliography}
\end{document}